\documentclass[preprintnumbers,amsmath,amssymb,twocolumn,nofootinbib,superscriptaddress,fleqn,showkeys,showpacs]{revtex4-1}
\usepackage[normalem]{ulem}
\usepackage{graphicx}
\usepackage{dcolumn}
\usepackage{bm}
\usepackage{color}
\usepackage[colorlinks,citecolor=blue,urlcolor=blue,linkcolor=blue]{hyperref}%??????¡Á????????
\newcommand{\n}{\nonumber}

\begin{document}
\title{Black ring entropy from the Weyl tensor}

\author{Ze-Wei Zhao}
\email{E-mail: zhaozewei@stumail.neu.edu.cn}
\affiliation{Department of Physics, College of Sciences, Northeastern University, Shenyang 110819, China}
\author{Chun-Kai Yu}
\email{E-mail: ckyu@mail.ustc.edu.cn}
\affiliation{Department of Physics, College of Sciences, Northeastern University, Shenyang 110819, China}
\affiliation{Department of Modern Physics, University of Science and Technology of China, Hefei 230026, China}
\author{Nan Li}
\email{E-mail: linan@mail.neu.edu.cn. Corresponding author.}
\affiliation{Department of Physics, College of Sciences, Northeastern University, Shenyang 110819, China}
\date{Received: date / Accepted: date}

\begin{abstract}
A black ring is an asymptotically flat vacuum solution of the $n$-dimensional Einstein equations with an event horizon of topology $S^1\times S^{n-3}$. In this study, a connection between the black ring entropy and the Weyl tensor $C_{\mu\nu\lambda\rho}$ is explored by interpreting the Weyl scalar invariant $C_{\mu\nu\lambda\rho} C^{\mu\nu\lambda\rho}$ as the entropy density in five-dimensional space-time. It is shown that the proper volume integral of $C_{\mu\nu\lambda\rho} C^{\mu\nu\lambda\rho}$ for a neutral black ring is proportional to the black ring entropy in the thin-ring limit. Similar calculations are extended to more general cases: a black string, a black ring with two angular momenta, and a black ring with a cosmological constant. The proportionality is also found to be valid for these complex black objects at the leading order.
\end{abstract}



\maketitle

\section{Introduction} \label{sec:intro}

Several decades ago, Penrose conjectured \cite{Penrose} that there could be some latent relationship between the Weyl tensor and the entropy of the gravitational field. This conjecture arose in cosmology on the basis of the fact that the universe evolves from an almost homogeneous and isotropic space-time to a phase with large-scale structures, during which the Weyl tensor, owing to its conformal symmetry, grows monotonically relative to the Ricci tensor. This monotonically increasing behavior reminds us of the second law of thermodynamics, and some scalar invariant constructed from the Weyl tensor could thus be identified with the entropy of the gravitational field. However, to our knowledge, the Penrose conjecture has not been physically interpreted or mathematically formulated in a rigorous way. It is not difficult to comprehend the obstacle, as entropy itself is essentially not a scalar, so a precise definition or even a specified notion of gravitational entropy via a scalar invariant seems impossible. Some early works on the dynamic and thermodynamic aspects of the Penrose conjecture can be found in \cite{Wainwright,Goode, Bonnor,Bonnor2, Husain,Goode2,Goode3, Newman,Hayward, Nesteruk,Rothman,Barve,Pelavas, Gron,Barrow,Lim,Rudjord, Li,Stoica,Maki,Li2,Marozzi,Okon}, and were summarized and reviewed in \cite{Ellis}.

However, it is still very plausible that the Weyl tensor could be related to the thermodynamics of the gravitational field and that some entropy measure could be constructed therefrom. This idea can be understood as follows. In mathematics, the curvature of an $n$-dimensional pseudo-Riemann manifold is described by the Riemann tensor $R_{\mu\nu\lambda\rho}$, which can be decomposed into the Ricci and Weyl sectors,
\begin{align}
R_{\mu\nu\lambda\rho}&=\frac{2}{n-2}(g_{\mu[\lambda}R_{\rho]\nu}-g_{\nu[\lambda}R_{\rho]\mu}) \n\\
&\quad-\frac{2}{(n-1)(n-2)}g_{\mu[\rho}g_{\lambda]\nu}R+C_{\mu\nu\lambda\rho}, \label{weyl define}
\end{align}
where $R_{\mu\nu}$ is the Ricci tensor, $R$ is the Ricci scalar, and $C_{\mu\nu\lambda\rho}$ is the Weyl tensor. However, in physics, the Einstein equations link the Ricci tensor only to the energy--momentum tensor of the gravitational field. As a result, the Weyl tensor is absent in classical general relativity; in other words, it is locally independent of the energy--momentum tensor. Therefore, we may infer that the information encoded in the Weyl tensor is associated not with the dynamic but with the thermodynamic aspects of gravitation. This observation provides a new and purely geometrical perspective on the thermodynamics of the gravitational field and can serve as a physical realization of the Penrose conjecture in a first step. Because the Weyl tensor is traceless and shares the same symmetries as the Riemann tensor, the Weyl scalar invariant (i.e., its complete contraction),
\begin{align}
W:=C_{\mu\nu\lambda\rho}C^{\mu\nu\lambda\rho}, \n
\end{align}
is the principal scalar invariant that we can construct. Actually, $W$ satisfies the requirements for defining the entropy from the Weyl tensor, as claimed in \cite{Ellis}: measure the local anisotropy of the free gravitational field, increase monotonically in the universe, and reproduce the black hole entropy formula. In \cite{Barve2}, the Weyl scalar invariant was even consulted to justify the cosmic censorship hypothesis. In this paper, we will focus on the relationship between the Weyl scalar invariant and the black ring entropy.

Below, we explain our idea more quantitatively. We first show how the Weyl tensor can be related to the black hole entropy. The metric of the Schwarzschild black hole is ${\rm d}s^2=-(1-R_{\rm S}/r)\,{\rm d}t^2+(1-R_{\rm S}/r)^{-1}\,{\rm d}r^2+r^2\,{\rm d}\theta^2 +r^2\sin^2\theta\,{\rm d}\phi^2$, where the event horizon is located at the Schwarzschild radius $R_{\rm S}=2GM$, and $M$ is the black hole mass. The Weyl scalar invariant is  obtained directly as
\begin{align}
W=\frac{12R_{\rm S}^2}{r^6}. \n
\end{align}
On the other hand, the Bekenstein--Hawking formula \cite{B--H1,B--H2} indicates that the entropy of the Schwarzschild black hole is
\begin{align}
S=\frac{A}{4G}=\frac{\pi R_{\rm S}^2}{G}, \n
\end{align}
where $A=4\pi R_{\rm S}^2$ is the event horizon area. Comparing these two results, we immediately observe the proportionality between $W$ and $S$. This proportionality inspires us to go further. Because the Weyl scalar invariant depends on the radial coordinate $r$ and does not decrease inside the event horizon, if it is evaluated along the world-lines of physical observers, we may make an educated interpretation that the Weyl scalar invariant $W$ is, or at least is proportional to, the entropy density of the gravitational field of the Schwarzschild black hole, and the Bekenstein--Hawking entropy could thus be obtained by taking its proper volume integral.

However, a dimensional analysis invalidates this simple attempt. In four-dimensional space-time, the dimension of the Weyl scalar invariant, $[W]$, is $+4$ in the natural system of units, but the dimension of the proper volume element, $[{\rm d}V_3]$, is $-3$. Therefore, the volume integral $\int W\,{\rm d}V_3$ has the dimension $+1$, but entropy is dimensionless. Consequently, we may regard the Weyl scalar invariant as the entropy density only in five-dimensional space-time, in which $[{\rm d}V_4]=-4$, so the proper volume integral,
\begin{align}
\int W\,{\rm d}V_4, \label{3}
\end{align}
is dimensionless and could lead to the correct entropy formula. We may generally consider this problem in an $n$-dimensional space-time, in which $[{\rm d}V_{n-1}]=-(n-1)$, whereas $[W]$ remains $+4$. Hence, dimension 5 is the unique possibility that $W$ could be interpreted as the entropy density of the gravitational field.

In \cite{song}, this idea was first explored for the five-dimensional Schwarzschild and Schwarzschild--anti-de Sitter black holes. It was found that the integral in (\ref{3}) does lead to the Bekenstein--Hawking formula,
\begin{align}
S\propto\int W\,{\rm d}V_4, \n
\end{align}
where the coefficient of proportionality is $1/96$ for the Schwarzschild black hole and is $\Lambda$-dependent for the Schwarzschild--anti-de Sitter black hole, with $\Lambda$ being the cosmological constant.

The present study is a successor to the work of \cite{song}, as dimension 5 is indeed very special in high-dimensional general relativity and hence deserves more careful study. For instance, it is the only possibility, other than dimension 4, that allows globally asymptotically flat solutions of the Einstein equations. Therefore, in addition to the five-dimensional black holes with an ordinary spherical horizon topology studied in \cite{song}, we proceed in this study to consider another important and interesting family of black holes with nonspherical horizon topologies: the five-dimensional black ring (with one or two angular momenta) and black string. Detailed calculations show that the proper volume integral of the Weyl scalar invariant is also proportional to the entropy of the black ring or black string. In general, the coefficient of proportionality depends on the shape of the black ring, and in the limit of a thin black ring, the proportionality can be greatly simplified.

This paper is organized as follows. In Sect. \ref{sec:ring}, a five-dimensional ring coordinate system is introduced, in which the black ring metric can be expressed more conveniently. Next, we explore the possibility of obtaining the black ring entropy by integrating the Weyl scalar invariant in Sect. \ref{sec:entropy}. In Sect. \ref{sec:dis}, we further discuss this calculation for a black string, a black ring with two angular momenta, and a black ring with a cosmological constant. We present our conclusions in Sect. \ref{sec:con}. We work in the natural system of units and set $c=\hbar=k_{\rm B}=1$, but we still keep the five-dimensional gravitational constant $G_5$, for a reason that will become clear later.

\section{Black ring} \label{sec:ring}

In this section, we briefly list the relevant properties of a black ring to provide necessary background for the following calculations. A five-dimensional black ring is an asymptotically flat black hole solution with a regular event horizon of topology $S^1\times S^2$ \cite{Emparan:2001wn,Elvang, BR,BR2, Elvang:2006dd, Elvang:2007,Emparan2007z,Emparan2008z,two2} and is thus different from other five-dimensional black holes with spherical event horizons (e.g., the Myers--Perry black hole \cite{MP}). Therefore, the black ring is of great importance and interest in general relativity, as it serves as explicit evidence indicating the invalidity of the uniqueness theorems in high-dimensional space-times (see \cite{BR2,Emparan2007z, Emparan2008z,two2} for excellent reviews). Unlike a black string or black brane, a black ring is curled, and its total length is finite. This makes the black ring more stable \cite{Elvang:2006dd} under small perturbations \cite{Gregory:1993vy,Gregory:1994bj}. To avoid gravitational collapse, a black ring must rotate or be charged in order to counteract the contraction. In five-dimensional space-time, a black ring may rotate in two mutually orthogonal planes and thus have two independent angular momenta. In the following, we investigate neutral black rings with one angular momentum (Sect. \ref{sec:entropy}) and with two (Sect. \ref{sec:two}).

We start from a so-called ring coordinate system \cite{BR2}, which makes it easier to comprehend the geometry of the black ring. To see it clearly, let us first consider a four-dimensional Euclidean space and choose polar coordinates for two orthogonal planes, so the metric is
\begin{align}
{\rm d}s_4^2={\rm d}r_1^2+r_1^2\,{\rm d}\phi^2+{\rm d}r_2^2+r_2^2\,{\rm d}\psi^2,\label{flat}
\end{align}
where $r_1$, $\phi$ and $r_2$, $\psi$ denote the radial and angular coordinates, respectively, in the two planes. Now, in the ring coordinate system, the coordinate transformation and inverse transformation are
\begin{align}
x&=-\frac{r_1^2+r_2^2-R^2}{\Sigma},\quad y=-\frac{r_1^2+r_2^2+R^2}{\Sigma} \label{bianhuan}
\end{align}
and
\begin{align}
r_1&=\frac{\sqrt{1-x^2}}{x-y}R, \quad r_2=\frac{\sqrt{y^2-1}}{x-y}R, \label{xy}
\end{align}
respectively, where $\Sigma=\sqrt{(r_1^2+r_2^2+R^2)^2-4r_2^2R^2}$, and $R$ characterizes the scale of the black ring ($\phi$ and $\psi$ are unchanged). From (\ref{xy}), it is easy to find the coordinate ranges $-1\leq x\leq1$ and $-\infty\leq y\leq -1$. Therefore, the coordinate $r_1=0$ or $r_2=R$ corresponds to $y=-\infty$, and asymptotic infinity is redefined as $x\to-1$ and $y\to -1$. A visualization in \cite{BR2} explains the geometric meaning of the ring coordinate system. In particular, the $y$ coordinate represents the distance from the black ring to the ring-shaped constant-$y$ hypersurface. With these transformations, the four-dimensional Euclidean metric in (\ref{flat}) can be expressed in terms of the ring coordinates:
\begin{widetext}
\begin{align}
{\rm d}s_4^2=\frac{R^2}{(x-y)^2}\left[\frac{{\rm d}x^2}{1-x^2}+(1-x^2)\,{\rm d}\phi^2-\frac{{\rm d}y^2}{1-y^2}-(1-y^2)\,{\rm d}\psi^2\right].\label{flat2}
\end{align}

We now place a circular neutral black ring at $r_1=0$ and $r_2=R$, and let it rotate only along the $\psi$ direction. The metric of the five-dimensional black ring preserves most of the properties in (\ref{flat2}) and reads \cite{Emparan:2001wn,two2}
\begin{align}
{\rm d}s^2=-\frac{F(y)}{F(x)}\left({\rm d}t-\sqrt{\frac{\lambda(\lambda-\nu)(1+\lambda)}{1-\lambda}}\frac{1+y}{F(y)}R\,{\rm d}\psi\right)^2
+\frac{R^2F(x)} {(x-y)^2}\left[\frac{{\rm d}x^2}{G(x)}+\frac{G(x)}{F(x)}\,{\rm d}\phi^2-\frac{{\rm d}y^2}{G(y)}-\frac{G(y)}{F(y)}\,{\rm d}\psi^2\right], \label{metric}
\end{align}
\end{widetext}
where $F(\xi)=1+\lambda\xi$, and $G(\xi)=(1-\xi^2)(1+\nu\xi)$. Two dimensionless parameters, $\lambda$ and $\nu$, reflect the rotation velocity and shape of the black ring (smaller values of $\nu$ correspond to thinner rings), and they vary within the range $0<\nu\leq\lambda<1$. Naturally, we can recover the four-dimensional Euclidean metric in (\ref{flat2}) in the limit $\lambda=\nu\to 0$. Moreover, to avoid the conical singularities at $x=-1$ and $y=-1$, $\phi$ and $\psi$ must be identified with the periods
\begin{align}
\Delta \phi=\Delta \psi=2\pi\frac{\sqrt{1-\lambda}}{1-\nu}.\label{psi}
\end{align}
Furthermore, to avoid another conical singularity at $x=1$, we must also set $\Delta \phi=2\pi\sqrt{1+\lambda}/(1+\nu)$, so the relation between $\lambda$ and $\nu$ is
\begin{align}
\lambda=\frac{2\nu}{1+\nu^2}.\label{lamb}
\end{align}
This relation is termed the equilibrium condition. With this restriction, (\ref{psi}) can finally be reduced to
\begin{align}
\Delta \phi=\Delta \psi =\frac{2\pi}{\sqrt{1+\nu^2}}.\n
\end{align}
Altogether, there remain only two independent parameters in (\ref{metric}), $R$ and $\nu$, which represent the scale and shape of the black ring, respectively. The angular momentum must be tuned to balance the gravitational self-attraction, so there is no longer a free parameter.

\section{Black ring entropy from the Weyl scalar invariant} \label{sec:entropy}

In the following, we calculate in detail the proper volume integral of the Weyl scalar invariant for a black ring and show that it is proportional to the black ring entropy.

From (\ref{weyl define}), the explicit form of the Weyl tensor in five-dimensional space-time reads
\begin{align}
C_{\mu\nu\lambda\rho}&=R_{\mu\nu\lambda\rho}+\frac{1}{3}(g_{\nu\lambda}R_{\mu\rho}+g_{\mu\rho}R_{\nu\lambda}-g_{\mu\lambda}R_{\nu\rho} \n\\
&\quad -g_{\nu\rho}R_{\mu\lambda})+\frac{1}{12}(g_{\nu\rho}g_{\mu\lambda}-g_{\nu\lambda}g_{\mu\rho})R.\label{weyl}
\end{align}
Hence, from (\ref{metric}) and (\ref{weyl}), it is straightforward to obtain the Weyl scalar invariant $W$ for the black ring. The complete expression (a polynomial with respect to the $y$ coordinate) is mathematically rather tedious, but fortunately, only its leading-order term is physically relevant, because the dominant contribution of $W$ to the integral in (\ref{3}) comes from the region around the black ring, where $y$ tends to $-\infty$. Even far from this region, the absolute value of $y$ is still much larger than 1 in the ring coordinate system, so we are always allowed to consider only the leading order term of $W$. Thus,
\begin{align}
W=\frac{12(1+\nu^2)^2\nu^2y^6}{(1+2x\nu+\nu^2)^2R^4}.\label{ring CC}
\end{align}
Furthermore, the leading-order term (with respect to $y$) of the four-dimensional proper volume element is
\begin{align}
{\rm d}V_4=\frac{(1+\nu)^{3/2}\sqrt{1+2x\nu+\nu^2}R^4}{(1+\nu^2)\sqrt{1-\nu}y^4}\,{\rm d}x{\rm d}\phi{\rm d}y{\rm d}\psi.\label{volume}
\end{align}
The horizon-crossing problem will be carefully discussed at the end of this section.

From (\ref{ring CC}) and (\ref{volume}), the proper volume integral of the Weyl scalar invariant is
\begin{align}
\int W\,{\rm d}V_4=\int\frac{12(1+\nu^2)(1+\nu)^{3/2}\nu^2y^2}{\sqrt{1-\nu}(1+2x\nu+\nu^2)^{3/2}}\,{\rm d}x{\rm d}\phi{\rm d}y{\rm d}\psi. \n
\end{align}
We may first safely perform the integrals for the $x$, $\phi$, and $\psi$ coordinates, where the integral intervals are $-1\leq x\leq 1$, $0\leq\phi\leq 2\pi/\sqrt{1+\nu^2}$, and $0\leq\psi\leq 2\pi/\sqrt{1+\nu^2}$, respectively. As a result,
\begin{align}
\int W\,{\rm d}V_4=\int \frac{96\pi^2\sqrt{1+\nu}\nu^2y^2}{(1-\nu)^{3/2}}\,{\rm d}y.\n
\end{align}
However, for the $y$ coordinate, we cannot directly set the integral interval to $-\infty\leq y\leq -1$, as the lower limit $-\infty$ will blow up the integral. Consequently, we first integrate $y$ from a lower limit, $-y_0$ ($y_0\gg 1$), to $-1$ (i.e., from the region just slightly off the black ring to asymptotic infinity) and then discuss the physical interpretation of $y_0$, so as to obtain a meaningful result. Thus,
\begin{align}
\int W\,{\rm d}V_4&=\int ^{-1}_{-y_0} \frac{96\pi^2 \sqrt{1+\nu}\nu^2y^2}{(1-\nu)^{3/2}}\,{\rm d}y \n\\
&=\frac{32\pi^2\sqrt{1+\nu}\nu^2y_0^3}{(1-\nu)^{3/2}}, \label{S0}
\end{align}
where the fact that $y_0\gg 1$ has been taken into account.

We now discuss the physical meaning of the lower limit $-y_0$. From (\ref{bianhuan}), setting $r_1=0$ yields
\begin{align}
y=-\frac{r_2^2+R^2}{|r_2^2-R^2|},\n
\end{align}
and $y=-\infty$ corresponds to $r_2=R$. However, it is generally acknowledged that classical general relativity is invalid at very small scales (e.g., several Planck lengths), where significant quantum gravity effects emerge. Hence, we may choose the lower limit $-y_0$ in such a way that the corresponding ring coordinate $r_2$ satisfies
\begin{align}
r_2=R\pm l_5, \label{l5}
\end{align}
where $l_5=\sqrt[3]{G_5}$ is the five-dimensional Planck length. Therefore,
\begin{align}
-y_0=-\frac{(R\pm l_5)^2+R^2}{|(R\pm l_5)^2-R^2|}\approx-\frac{R}{l_5},\n
\end{align}
where we have used the fact that $R\gg l_5$, as the scale of the black ring should certainly be much larger than the Planck length. If we instead set $r_2=R$ initially and choose the lower limit $-y_0$ such that $r_1=\pm l_5$, we will arrive at the same result that $-y_0\approx-{R}/{l_5}$.

With this choice, (\ref{S0}) becomes
\begin{align}
\int W\,{\rm d}V_4=\frac{32\pi^2\sqrt{1+\nu}\nu^2R^3}{(1-\nu)^{3/2}l_5^3}. \label{S1}
\end{align}
This is the final result for the proper volume integral of the Weyl scalar invariant. Here, we should point out that our strategy to obtain (\ref{S1}) is still semiclassical. The choice in (\ref{l5}) is equivalent to introducing a cutoff for the $y$ coordinate to avoid the intrinsic singularity at the black ring, and in this way, we eventually arrive at a physically meaningful result in (\ref{S1}).

Further, the Bekenstein--Hawking entropy of the black ring can be obtained by evaluating its event horizon area. From (\ref{metric}), the black ring has a regular event horizon at $y_{\rm h}=-1/\nu$, so an easy integral over the $x$, $\phi$, and $\psi$ coordinates yields its area,
\begin{align}
A=\frac{8\sqrt{2}\pi^2\nu^2R^3}{(1-\nu)(1+\nu^2)^{3/2}}.\n
\end{align}
Thus, the black ring entropy is simply written as
\begin{align}
S=\frac{A}{4G_5}=\frac{A}{4l_5^3}=\frac{2\sqrt{2}\pi^2\nu^2R^3}{(1-\nu)(1+\nu^2)^{3/2}l_5^3}.\label{S2}
\end{align}

So far, from (\ref{S1}) and (\ref{S2}), we can clearly observe the proportionality between the black ring entropy and the proper volume integral of the Weyl scalar invariant (both of which are proportional to $ R^3/l_5^3$). The coefficient of proportionality generally depends on the shape parameter $\nu$ of the black ring. This complexity may be largely eliminated if we focus on thin black rings with $\nu\ll 1$. In other words, if we consider only the leading-order terms (with respect to $\nu$) in (\ref{S1}) and (\ref{S2}),
\begin{align}
\int W\,{\rm d}V_4=\frac{32\pi^2\nu^2R^3}{l_5^3}, \label{SC}
\end{align}
and
\begin{align}
S=\frac{2\sqrt{2}\pi^2\nu^2R^3}{l_5^3}, \label{zhao}
\end{align}
we immediately arrive at
\begin{align}
S=\frac{\sqrt{2}}{16}\int W\,{\rm d}V_4.\label{lll}
\end{align}
The above result supports our idea that the Weyl scalar invariant can be regarded as the entropy density of the gravitational field in five-dimensional space-time, and its proper volume integral thus leads to the black ring entropy. This idea was first explored in \cite{song} for the five-dimensional Schwarzschild and Schwarzschild--anti-de Sitter black holes with a simple spherical horizon topology and is now investigated again for the more complex five-dimensional black ring with a nonspherical horizon topology. Actually, it will also be applied to other more general black ring solutions in the subsequent section.

Here, we should admit that the coefficient of proportionality for thin black rings, $\sqrt{2}/16$, seems different from that for the Schwarzschild black hole, $1/96$ \cite{song}. This is because the choice of the ring coordinate $r_2$ in (\ref{l5}) is just a preliminary estimate. We may set $r_2=R\pm al_5$ for generality, where $a$ is some number in ${\cal O}(1)$, and adjusting $a$ may make things look better. However, the explicit value of $a$ is not highly relevant, and what we are really concerned with is the proportionality between the volume integral and the entropy.

Finally, we should discuss the horizon-crossing problem more carefully when integrating the $y$ coordinate. Strictly speaking, the leading-order term of the four-dimensional proper volume element should be
\begin{align}
{\rm d}V_4=\frac{(1+\nu)^{3/2}\sqrt{(1+2x\nu+\nu^2)\nu|y|}R^4}{(1+\nu^2)\sqrt{(1-\nu)|1+\nu y|}y^4}\,{\rm d}x{\rm d}\phi{\rm d}y{\rm d}\psi.\n
\end{align}
Naturally, this result reduces to that in (\ref{volume}) if $y\ll-1$. However, when $y$ crosses the event horizon at $y_{\rm h}=-1/\nu$, the term
$1+\nu y$ in the denominator will change sign. Therefore, strictly speaking, the proper volume integral of the Weyl scalar invariant should be
\begin{align}
\int W\,{\rm d}V_4&=\frac{96 \pi^2\sqrt{(1+\nu)\nu^5}}{(1-\nu^2)^{3/2}}\times \n\\
&\quad\left(\int^{y_{\rm h}}_{-y_0}\sqrt{\frac{y^5}{1+\nu y}}\,{\rm d}y+\int^{-1}_{y_{\rm h}}\sqrt{\frac{-y^5}{1+\nu y}}\,{\rm d}y\right).\n
\end{align}
The first integral will give exactly the same result as (\ref{S0}) in the limit $y_0\gg 1$, meaning that only the region around the black ring dominates the integral. The second integral (i.e., the contribution outside the event horizon) is a trivial number independent of $y_0$ and can thus be safely disregarded. These observations validate our assessment that there is effectively no horizon-crossing problem in the volume integral.

\section{Discussion} \label{sec:dis}

In this section, we discuss some more complicated types of black rings: a black string, a black ring with two angular momenta, and a black ring with a cosmological constant. The corresponding calculations support our result in the simplest case of a black ring with one angular momentum in Sect. \ref{sec:entropy}.

\subsection{Black string} \label{sec:string}

A black string can be viewed as an extreme black ring with an infinite scale $R$. Therefore, we may obtain the metric of a boosted black string directly from (\ref{metric}) in the limit $R\to\infty$. First, it is convenient to transform the coordinates from $x$, $y$, and $\psi$ in (\ref{metric}) to $r$, $\theta$, and $z$,
\begin{align}
r=-\frac Ry, \quad \cos\theta=x, \quad z=R\psi, \n
\end{align}
and extend the string along the $z$ direction ($\phi$ is unchanged). The parameters $\lambda$ and $\nu$ in (\ref{metric}) may also be reparameterized in terms of $r_0$ and $\sigma$:
\begin{align}
\lambda=\frac{r_0\cosh^2\sigma}{R}, \quad \nu=\frac{r_0}{R}. \label{sigma}
\end{align}
If we take the limit $R\to\infty$ while keeping the coordinates $r$, $\theta$, and $z$ and the parameters $r_0$ and $\sigma$ finite, (\ref{metric}) becomes
\begin{align}
{\rm d}s^2&=-\widehat{f}\left({\rm d}t+\frac{r_0}{r\widehat{f}} \sinh\sigma \cosh\sigma\,{\rm d}z\right)^2+\frac{f}{\widehat{f}}\,{\rm d}z^2 \n\\
&\quad+\frac{{\rm d}r^2}{f}+r^2\,{\rm d}\theta^2+r^2\sin^2\theta\,{\rm d}\phi^2,\label{string}
\end{align}
where $f=1-r_0/r$, and $\widehat{f}=1-r_0\cosh^2\sigma/r$. From (\ref{string}), it is easy to see that the event horizon of the black string is located at $r=r_0$. The equilibrium condition in (\ref{lamb}) is now expressed as $\cosh^2\sigma=2R^2/(R^2+r_0^2)$, and when $R\to\infty$,
\begin{align}
\cosh^2\sigma\to 2.\label{equili}
\end{align}
This is the limiting condition for maintaining the equilibrium of a boosted black string of very large scale $R$.

From (\ref{weyl}) and (\ref{string}), we obtain the Weyl scalar invariant for the black string,
\begin{align}
W=\frac{12r_0^2}{r^6},\n
\end{align}
and the four-dimensional proper volume element is
\begin{align}
{\rm d}V_4=r^2 \sin\theta \sqrt{\left|\frac{2r-r_0+r_0\cosh{2\sigma}}{2r-2r_0}\right|}\,{\rm d}\theta{\rm d}\phi{\rm d}r{\rm d}z.\n
\end{align}

Before performing the integration, we should clarify the integral intervals for the four coordinates. First, from $\cos\theta=x\in[-1,1]$, we have $0\leq\theta\leq\pi$. Second, in the limit $R\to\infty$, the period of $\phi$ is simplified to $0\leq\phi\leq2\pi$. Third, from $r=-R/y$, $-R/l_5\leq y\leq -1$ corresponds to $l_5\leq r\leq R$ ($R\to \infty$). Finally, we should set $0\leq z\leq\Delta z$, meaning that we integrate over only a section of the black string. Following the same procedure as in Sect. \ref{sec:entropy}, we arrive at the proper volume integral of the Weyl scalar invariant for the boosted black string, 
\begin{align}
\int W\,{\rm d}V_4 =\frac{16\pi r_0^2\Delta z}{l_5^3},\label{weyl string}
\end{align}
where we have taken into account the equilibrium condition in (\ref{equili}) and the fact that the integral is dominated only by the region where $r\ll r_0$, which greatly simplifies the calculation.

Furthermore, the event horizon area of the section of the black string is obtained by integrating the area element at the horizon, where $r=r_0$. From (\ref{string}) and (\ref{equili}), we have
\begin{align}
A=\int\sqrt{2}r_0^2\sin\theta\,{\rm d}\theta{\rm d}\phi{\rm d}z=4\sqrt{2}\pi r_0^2\Delta z.\n
\end{align}
Therefore, the black string entropy is
\begin{align}
S=\frac{A}{4G_5}=\frac{\sqrt{2}\pi r_0^2\Delta z}{l_5^3}.\label{string entropy}
\end{align}

From (\ref{weyl string}) and (\ref{string entropy}), we again obtain the proportionality between the proper volume integral of the Weyl scalar invariant and the black string entropy, with the same coefficient of proportionality as the black ring,
\begin{align}
S= \frac{\sqrt{2}}{16} \int W\,{\rm d}V_4.\n
\end{align}
This result is expected. Because the boosted black string, by its nature ($R\to \infty$ and $\nu=r_0/R\to 0$), can be viewed as an extremely thin black ring, they should share the same result. Moreover, if we take the length of the section of the black string as $\Delta z=2\pi R$, with the parameter transformation $\nu=r_0/R$ in (\ref{sigma}), we find that the black string entropy in (\ref{string entropy}) becomes the black ring entropy in (\ref{zhao}).

\subsection{Black ring with two angular momenta} \label{sec:two}

We move on to the entropy of a black ring with two independent angular momenta (i.e., a black ring with rotation not only along the $\psi$ direction of topology $S^1$ but also along the $\phi$ direction of topology $S^2$). The metric of such a black ring is a generalization of (\ref{metric}) physically but is much more complicated mathematically. It was first obtained in \cite{two} by using the complete integrability of the system via the inverse scattering method. The following expression is adopted from \cite{two2}:
\begin{widetext}
\begin{align}
{\rm d}s^2=-\frac{F(y,x)}{F(x,y)}({\rm d}t+\omega)^2+\frac{\widetilde{R}^2F(x,y)}{(x-y)^2(1-\alpha)^2}\left[\frac{{\rm d}x^2}{G(x)}-\frac{{\rm d}y^2}{G(y)}\right] +\frac{H(y,x)}{F(y,x)}\,{\rm d}\widetilde{\phi}^2-\frac{H(x,y)}{F(y,x)}\,{\rm d}\widetilde{\psi}^2-2\frac{J(x,y)}{F(y,x)}\,{\rm d}\widetilde{\phi}{\rm d}\widetilde{\psi}, \label{metric two}
\end{align}
where the coordinates $x$ and $y$ maintain essentially the same meanings as those in (\ref{metric}), the canonical periods of the angles $\widetilde {\phi}$ and $\widetilde{\psi}$ have been rescaled to $2\pi$, and the parameter $\widetilde{R}$ is related to $R$ in (\ref{metric}) as $\widetilde{R}^2= R^2/(1+\nu^2)$. If the black ring is in equilibrium (i.e., without conical singularities), the 1-form $\omega$ describing the rotations is 
\begin{align}
\omega=-\dfrac{\sqrt{2}\widetilde{R}\nu\sqrt{(1+\alpha)^2-\nu^2}}{F(y,x)}\left\{(1-x^2)y\sqrt{\alpha}\,{\rm d}\widetilde{\phi}+\dfrac{1+y}{1-\nu+\alpha}
\left[1+\nu-\alpha+x^2y\alpha(1-\nu-\alpha)+2\alpha x(1-y)\right]{\rm d}\widetilde{\psi}\right\},\n
\end{align}
and
\begin{align}
G(x)&=(1-x^2)(1+\nu x+\alpha x^2),\n\\
F(x,y)&=1+\nu^2-\alpha^2+2\nu\alpha(1-x^2)y+2x\nu(1-y^2\alpha^2)+x^2y^2\alpha(1-\nu^2-\alpha^2),\n\\
J(x,y)&=\frac{\widetilde{R}^2(1-x^2)(1-y^2)\nu\sqrt{\alpha}}{(x-y)(1-\alpha)^2}
\left[1+\nu^2-\alpha^2+2(x+y)\nu\alpha-xy\alpha(1-\nu^2-\alpha^2)\right], \n\\
H(x,y)&=\frac{\widetilde{R}^2}{(x-y)^2(1-\alpha)^2}\left(G(x)(1-y^2)\left\{[(1-\alpha)^2-\nu^2](1+\alpha)
+y\nu(1-\nu^2+2\alpha-3\alpha^2)\right\}\right.\n\\
& \quad+G(y)\left\{2\nu^2 +x\nu[(1-\alpha)^2+\nu^2]+x^2[(1-\alpha)^2-\nu^2](1+\alpha)\right. \n\\
& \quad\left.\left. +x^3\nu(1-\nu^2-3\alpha^2+2\alpha^3)-x^4(1-\alpha)\alpha(-1+\nu^2+\alpha^2)\right\}\right).\n
\end{align}
\end{widetext}
The second angular momentum is characterized by a new parameter, $\alpha$, the range of which is restricted to $0\leq\alpha<1$ and $2\sqrt{\alpha}\leq\nu<1+\alpha$. Obviously, by setting $\alpha=0$, we recover the metric in (\ref{metric}). The location of the event horizon is the solution for vanishing $G(y)$ in (\ref{metric two}):
\begin{align}
y_{\rm h}=\frac{-\nu+\sqrt{\nu^2-4\alpha}}{2\alpha}.\label{loc of hori}
\end{align}
Requiring that the root should be real yields the upper bound for $\alpha$, $\alpha\leq \nu^2/4$. Furthermore, the event horizon area is
\begin{align}
A=\frac{8\sqrt{2}\pi^2(1+\nu+\alpha)\nu\widetilde{R}^3}{(y_{\rm h}^{-1}-y_{\rm h})(1-\alpha)^2}. \label{horizon two}
\end{align}

The following calculations are totally parallel to those in Sect. \ref{sec:entropy}. However, the complete results are extremely tedious and irrelevant, so we focus on two special cases: (1) $\alpha\ll\nu^2/4$, such that the second angular momentum has a much smaller effect than the first one; (2) $\alpha=\nu^2/4$, such that the two angular momenta are comparable, and the event horizon is thus degenerate.

In the first case, the contribution of $\alpha$ is considered to be a small perturbation to the black ring with one angular momentum. Because $\alpha\ll\nu^2$, we set $\alpha\sim{\cal O}(\nu^3)$, and all the perturbative calculations are performed up to the orders of $\alpha$ and $\nu^3$. Under these conditions, the leading-order terms of the proper volume integral of the Weyl scalar invariant and the black ring entropy remain the same as in (\ref{SC}) and (\ref{zhao}),
\begin{align}
\int W\,{\rm d}V_4=\frac{32\pi^2\nu^2R^3}{l_5^3}, \quad S=\frac{2\sqrt{2}\pi^2\nu^2R^3}{l_5^3}, \n
\end{align}
and so does the proportionality between them. These results are natural, as the contribution of $\alpha$ should be absent at the leading order in the perturbative approach. We will not show the complicated higher-order corrections, as detailed calculations indicate that these contributions appear from at least the order of $\alpha^2$ (i.e., $\nu^6$), and thus the effects of the second angular momentum are negligible at not only the leading order but also the next-to-leading order.

In the second case, with a degenerate event horizon, all the calculations are greatly simplified. At the leading order, we obtain
\begin{align}
\int W\,{\rm d}V_4 &= \frac{40\pi^2\nu^2R^3}{l_5^3}.\label{weyl dege}
\end{align}
Moreover, from (\ref{loc of hori}) and the degeneracy condition, $\alpha=\nu^2/4$, we have $y_{\rm h}=-2/\nu$, and the leading-order term of the event horizon area in (\ref{horizon two}) reduces to $A= 4\sqrt{2}\pi^2\nu^2R^3$. Hence, the black ring entropy is
\begin{align}
S=\frac{A}{4G_5}=\frac{\sqrt{2}\pi^2\nu^2R^3}{l_5^3}. \label{entropy dege}
\end{align}
Finally, from (\ref{weyl dege}) and (\ref{entropy dege}), we have
\begin{align}
S=\frac{\sqrt{2}}{40} \int W\,{\rm d}V_4.\n
\end{align}
One may wonder why the coefficient of proportionality, $\sqrt{2}/40$, is not the same as that of a black ring or black string, $\sqrt{2}/16$. This is understandable, as it is the shape parameter $\nu$ that has physical significance, and the coefficient of proportionality does not encode much information. In this way, we confirm our interpretation of the Weyl scalar invariant as the entropy density of a black ring with two angular momenta. In addition, we should keep in mind that what we study are still thin black rings.

\subsection{Black ring with a cosmological constant} \label{sec:dS}

Last, we briefly discuss a black ring with a cosmological constant. Unfortunately, the existence of such a black ring solution is still an unsolved problem. It is acknowledged that an asymptotically flat black ring solution is always obtained owing to the integrability of the Einstein equations. However, it is still unclear whether the cosmological constant destroys the integrability \cite{Figueras}. If it does, the black ring solution with a cosmological constant is not supposed to exist. As a result, there are only some numerical black ring solutions in the literature \cite{Caldarelli}. Note that the black ring studied in \cite{Caldarelli} by approximate methods is also thin and is consistent with our result. It was shown in \cite{Caldarelli} that in five-dimensional anti-de Sitter space-time with negative $\Lambda$, the event horizon area of the black ring is $A=8\pi^2\nu^2R^3\sqrt{2+3R^2/L^2}$, where $L\sim 1/\sqrt{-\Lambda}$ is the characteristic cosmological radius. Thus, the black ring entropy becomes
\begin{align}
S=\frac{A}{4G_5}=\frac{2\pi^2 \nu^2R^3}{l_5^3}\sqrt{2+\frac{3R^2}{L^2}}. \label{entropy ads}
\end{align}
This result is identical to the leading-order term of the entropy in (\ref{zhao}) in the limit $\Lambda\to 0$ or $L\to\infty$.

Because the exact black ring metric with a cosmological constant is unavailable, we cannot calculate the Weyl scalar invariant directly. However, as shown in \cite{song}, the volume integral $\int W\,{\rm d}V_4$ is not likely to be influenced significantly by $\Lambda$ because the contribution of $\Lambda$ to the integral appears only at large distances, where it is strongly suppressed by the Weyl scalar invariant, which decays even faster. This comparison also indicates that it is the Weyl scalar invariant $C_{\mu\nu\lambda\rho}C^{\mu\nu\lambda\rho}$, not the Kretschmann scalar invariant $R_{\mu\nu\lambda\rho} R^{\mu\nu\lambda\rho}$, that can be interpreted as the entropy density. Although these two scalar invariants coincide in the absence of $\Lambda$, the Kretschmann scalar invariant is significantly modified by $\Lambda$ at large distances and thus diverges the integral.

Comparing (\ref{SC}) and (\ref{entropy ads}), we have
\begin{align}
S=\frac{\sqrt{2+3R^2/L^2}}{16} \int W\,{\rm d}V_4.\n
\end{align}
In this way, the proper volume integral of the Weyl scalar invariant is still proportional to the black ring entropy, but the coefficient of proportionality is $\Lambda$-dependent.

Finally, we should also state that the thermodynamics of a black ring with a positive cosmological constant (i.e., in de Sitter space-time) is still a controversial and poorly defined issue \cite{Bousso,Lin,Stuchlik,Myung,Balasubramanian,Ghezelbash, Setare,Cardoso,Shankaranarayanan,Myung2,Myung3, Araujo,Bhattacharya,Kubiznak,Khuri}, and we omit the corresponding discussion here.

\section{Conclusion} \label{sec:con}

The latent relationship between gravitation and thermodynamics is a hot topic in modern theoretical physics. Penrose's conjecture of the Weyl curvature is one of the explorations of the role of the Weyl tensor $C_{\mu\nu\lambda\rho}$ in black hole thermodynamics and cosmology. It was preliminarily investigated in \cite{song} by looking at the relation between the Weyl scalar invariant $C_{\mu\nu\lambda\rho}C^{\mu\nu\lambda\rho}$ and the entropy of five-dimensional black holes. It was found that the Weyl scalar invariant can be interpreted as the entropy density, and its proper volume integrals can correctly lead to the Bekenstein--Hawking entropies of the Schwarzschild and Schwarzschild--anti-de Sitter black holes.

In this study, we generalize the basic idea in \cite{song} to another important and interesting asymptotically flat vacuum solution of the five-dimensional Einstein equations: the black ring solution. We perform the proper volume integral of the Weyl scalar invariant for a neutral black ring in the ring coordinate system and show that it is proportional to the black ring entropy. Although for the Bekenstein--Hawking formula, $S\propto R^3/l_5^3$ can be understood as a consequence of dimensional analysis, $\int W\,{\rm d}V_4\propto R^3/l_5^3$ is a new insight. Furthermore, we extend our calculation to three more complicated cases. (1) For a black string, the proportionality is the same as that of a black ring, as it can be viewed as a limiting case of a black ring with an infinite scale. (2) For a black ring with two angular momenta, two special cases are studied: i) the second angular momentum is much smaller than the first one, and the proportionality still holds at the leading order in the perturbative approach; ii) the two angular momenta are comparable, and the coefficient of proportionality is modified by the second angular momentum. (3) The black ring with a cosmological constant is not extensively discussed, as the black ring solution is not thought to exist owing to the lack of an integrability condition. However, the proportionality is still expected, as the calculation in \cite{song} indicated that the cosmological constant would not influence the proper volume integral significantly. In conclusion, these various calculations support our interpretation of the Weyl scalar invariant as the entropy density of the gravitational field.

Finally, we should also discuss some shortcomings of our work. First, we did not discuss the five-dimensional Myers--Perry black hole \cite{MP}, which is a natural generalization of the Kerr solution with a spherical horizon topology. For the Myers--Perry solution, there is a region where the value of its Weyl scalar invariant is negative, and our attempt to interpret it as the entropy density will thus be invalid. Although the volume integral can still be formally performed regardless of the sign of $W$, its physical meaning will be unclear. Therefore, we are concerned with only the black hole solutions whose Weyl scalar invariants are always positive. Second, the proper volume integral of the Weyl scalar invariant is divergent if it is performed in the entire four-dimensional space. This divergence is generally unavoidable owing to the intrinsic singularity at the black ring; in this study, we introduce an effective cutoff in the integration of the ring coordinate, as was done in \cite{song}. This means that we integrate only over the region where classical general relativity can safely be trusted, so our approach in calculating the black ring entropy is geometrical, not quantum. Third, the coefficient of proportionality between $S$ and $\int W\,{\rm d}V_4$ varies for different black objects (i.e., it depends on the shape parameter $\nu$ or the cosmological constant $\Lambda$). This limits the ability to calculate $S$ from $\int W\,{\rm d}V_4$, except under some limiting circumstances. For instance, for thin black rings, both $S$ and $\int W\,{\rm d}V_4$ are proportional to $\nu^2$, so their ratio is merely a number. Our approach should still be considered as a first qualitative step to understanding the Penrose conjecture, but this result is nontrivial and deserves further study. Finally, all our calculations for black holes and black rings are valid in five-dimensional space-time, but not the ordinary four-dimensional one. The next step is naturally to obtain the Bekenstein--Hawking formula for four-dimensional black holes from the Weyl tensor. This can be realized by imagining that the mass of a five-dimensional black ring is distributed along a compact extra dimension, and if this dimension is wrapped to an extremely small scale, the black ring may be effectively viewed as a black hole from a four-dimensional perspective. Some of our related calculations have supported this attempt, and it will be our next topic of research.

\begin{acknowledgements}
We are very grateful to Zhao-Hui Chen, Dominik J. Schwarz, Shu-Peng Song, and Liu Zhao for fruitful discussions. This work is supported by the Fundamental Research Funds for the Central Universities of China (No. N170504015).
\end{acknowledgements}

\end{document}